\documentclass[aps,prd,preprint,tightenlines,superscriptaddress,showpacs,byrevtex]{revtex4}

\usepackage{graphicx} % Include figure files
\usepackage{dcolumn}  % Align table columns on decimal point

\def\piz{\pi ^0 }
\def\pip{\pi ^+ }

\def\B0b{\bar{B}^{0}}
\def\BBbar{B\bar{B}}
\def\qqbar{q\bar{q}}

\def\epem{e^+ e^-}

\def\mpmm{\mu ^+ \mu ^-}

\def\Mmiss{M_{\rm miss}}
\def\Mm2{\Mmiss^2}

\def\ACostt{\left| \cos \theta _T \right|}
\def\GeV{{\rm GeV}}
\def\GeVc{{\rm GeV\!/}c}
\def\GeVcc{{\rm GeV\!/}c^2}
\def\MeV{{\rm MeV}}

\def\MeVcc{{\rm MeV\!/}c^2}
\def\ifb{{\rm fb^{-1}}}
\def\to{\rightarrow}
\def\BR{{\cal B}}
\def\pmuB{p_\mu^{B}}

\def\Btolv{B^+\to \ell^+ \nu_\ell}
\def\Btoev{B^+\to e^+ \nu_e}
\def\Btomuv{B^+\to \mu^+ \nu_\mu}
\def\Btolvg{B^+\to \ell^+ \nu_\ell \gamma}
\def\Btoevg{B^+\to e^+ \nu_e \gamma}
\def\Btomuvg{B^+\to \mu^+ \nu_\mu \gamma}

\def\evg{e^+ \nu_e \gamma}
\def\muvg{\mu^+ \nu_\mu \gamma}

\def\Ebeam{E_{\rm beam}}
\def\Mbc{M_{\rm bc}}
\def\DE{\Delta E}
\def\pLep{p_\ell}
\def\Xulnu{X_u \ell \nu_\ell}
\def\BtoXulnu{B\to X_u \ell \nu_\ell}

\def\ResultBRBtomuv{2.0 \times 10^{-6}}
\def\ResultBRBtoevg{2.2 \times 10^{-5}}
\def\ResultBRBtomuvg{2.3 \times 10^{-5}}
\graphicspath{{ps}}

\begin{document}

%\vspace*{-3\baselineskip}
%\resizebox{!}{3cm}{\includegraphics{belle.eps}}

\preprint{\vbox{ \hbox{BELLE-CONF-0429}
                 \hbox{ICHEP04 11-0676}
%                 \hbox{hep-ex nnnn, if available}
%                 \hbox{version 0.97}
}}

\title{ \quad\\[0.5cm]  Search for $\Btomuv$ and $\Btolvg$ decays}

%%%% >>>>> insert the authorlist here. BEFORE the abstract !!!!! <<<<<

%%% Paper:
%%% Journal:  summer 2004 conference papers (PRL format)
%%% Contacts:
%%% Last revised on July 14, 2004 16:40:00 EDT
%%% Non-responding authors or those who said NO are commented out.
%%% ====================================================================
%%% Click the RELOAD button on your web browser to see the updated file.
%%% ====================================================================
%%% Use \input{author} to insert this material into your latex file.
%%%%% Force institutions to appear in alphabetical order when typeset.
\affiliation{Aomori University, Aomori}
\affiliation{Budker Institute of Nuclear Physics, Novosibirsk}
\affiliation{Chiba University, Chiba}
\affiliation{Chonnam National University, Kwangju}
\affiliation{Chuo University, Tokyo}
\affiliation{University of Cincinnati, Cincinnati, Ohio 45221}
\affiliation{University of Frankfurt, Frankfurt}
\affiliation{Gyeongsang National University, Chinju}
\affiliation{University of Hawaii, Honolulu, Hawaii 96822}
\affiliation{High Energy Accelerator Research Organization (KEK), Tsukuba}
\affiliation{Hiroshima Institute of Technology, Hiroshima}
\affiliation{Institute of High Energy Physics, Chinese Academy of Sciences, Beijing}
\affiliation{Institute of High Energy Physics, Vienna}
\affiliation{Institute for Theoretical and Experimental Physics, Moscow}
\affiliation{J. Stefan Institute, Ljubljana}
\affiliation{Kanagawa University, Yokohama}
\affiliation{Korea University, Seoul}
\affiliation{Kyoto University, Kyoto}
\affiliation{Kyungpook National University, Taegu}
\affiliation{Swiss Federal Institute of Technology of Lausanne, EPFL, Lausanne}
\affiliation{University of Ljubljana, Ljubljana}
\affiliation{University of Maribor, Maribor}
\affiliation{University of Melbourne, Victoria}
\affiliation{Nagoya University, Nagoya}
\affiliation{Nara Women's University, Nara}
\affiliation{National Central University, Chung-li}
\affiliation{National Kaohsiung Normal University, Kaohsiung}
\affiliation{National United University, Miao Li}
\affiliation{Department of Physics, National Taiwan University, Taipei}
\affiliation{H. Niewodniczanski Institute of Nuclear Physics, Krakow}
\affiliation{Nihon Dental College, Niigata}
\affiliation{Niigata University, Niigata}
\affiliation{Osaka City University, Osaka}
\affiliation{Osaka University, Osaka}
\affiliation{Panjab University, Chandigarh}
\affiliation{Peking University, Beijing}
\affiliation{Princeton University, Princeton, New Jersey 08545}
\affiliation{RIKEN BNL Research Center, Upton, New York 11973}
\affiliation{Saga University, Saga}
\affiliation{University of Science and Technology of China, Hefei}
\affiliation{Seoul National University, Seoul}
\affiliation{Shinshu University, Matsumoto}
\affiliation{Sungkyunkwan University, Suwon}
\affiliation{University of Sydney, Sydney NSW}
\affiliation{Tata Institute of Fundamental Research, Bombay}
\affiliation{Toho University, Funabashi}
\affiliation{Tohoku Gakuin University, Tagajo}
\affiliation{Tohoku University, Sendai}
\affiliation{Department of Physics, University of Tokyo, Tokyo}
\affiliation{Tokyo Institute of Technology, Tokyo}
\affiliation{Tokyo Metropolitan University, Tokyo}
\affiliation{Tokyo University of Agriculture and Technology, Tokyo}
\affiliation{Toyama National College of Maritime Technology, Toyama}
\affiliation{University of Tsukuba, Tsukuba}
\affiliation{Utkal University, Bhubaneswer}
\affiliation{Virginia Polytechnic Institute and State University, Blacksburg, Virginia 24061}
\affiliation{Yonsei University, Seoul}
  \author{K.~Abe}\affiliation{High Energy Accelerator Research Organization (KEK), Tsukuba} % KEK
  \author{K.~Abe}\affiliation{Tohoku Gakuin University, Tagajo} % TohokuGakuin
  \author{N.~Abe}\affiliation{Tokyo Institute of Technology, Tokyo} % TIT
  \author{I.~Adachi}\affiliation{High Energy Accelerator Research Organization (KEK), Tsukuba} % KEK
  \author{H.~Aihara}\affiliation{Department of Physics, University of Tokyo, Tokyo} % Tokyo
  \author{M.~Akatsu}\affiliation{Nagoya University, Nagoya} % Nagoya
  \author{H.~An}\affiliation{Yonsei University, Seoul} % Yonsei
  \author{Y.~Asano}\affiliation{University of Tsukuba, Tsukuba} % Tsukuba
  \author{T.~Aso}\affiliation{Toyama National College of Maritime Technology, Toyama} % Toyama
  \author{V.~Aulchenko}\affiliation{Budker Institute of Nuclear Physics, Novosibirsk} % BINP
% ...
% Lausanne
\collaboration{The Belle Collaboration}

\begin{abstract}
We have searched for the leptonic and radiative leptonic $B$ decays,
$\Btomuv$, $\Btoevg$ and $\Btomuvg$.  Using a $140~\ifb$ data sample collected
with the Belle detector at the KEKB asymmetric $\epem$ collider, we
find no evidence for signals in any mode and set the following
preliminary upper limits at 90\% confidence level:
$\BR(\Btomuv) < \ResultBRBtomuv $,
$\BR(\Btoevg) < \ResultBRBtoevg $ and
$\BR(\Btomuvg) < \ResultBRBtomuvg $.
\end{abstract}


\maketitle

{\renewcommand{\thefootnote}{\fnsymbol{footnote}}}
\setcounter{footnote}{0}

\section{Introduction}

In the Standard Model (SM), the fully-leptonic decays $\Btolv$,
where $\ell$ represents an electron, muon, or $\tau$ lepton, are
allowed through annihilation into a virtual $W$ boson.  The
branching fraction is given by:
$$\BR (\Btolv) = \frac{G^2_F m_B m^2_\ell}{8\pi} \left( 1 -
\frac{m^2_\ell}{m^2_B} \right)^2 f_{B}^{2} |V_{ub}|^2 \tau_B\ , $$
where $G_F$ is the Fermi coupling constant, $m_B$ and $m_\ell$ are the
masses of the $B$ meson and lepton, $\tau_B$ is the $B$ meson
lifetime, $V_{ub}$ is an element of the Cabibbo-Kobayashi-Maskawa
quark mixing matrix~\cite{CKM}, and $f_B$ is the decay constant that
parameterizes the overlap of the quark wave functions within the
meson. A measurement of this branching fraction, combined with the
value of $|V_{ub}|$ obtained from other decay modes such as $ B
\rightarrow \pi \ell \bar{\nu_\ell} $, allows us to determine $f_B$,
which is needed to extract $|V_{td}|$ from measurements of
$B^0\bar{B^0} $ mixing.  The theoretical expectation for the branching
fraction of $B^+ \rightarrow \tau^+ \nu_\tau$ lies in the range \( ( 1
- 10 ) \times 10^{-5} \).  Since the branching fraction is
proportional to the square of the mass of the charged lepton, the
branching fractions for $\Btomuv$ and $\Btoev$ are suppressed by
factors of 225 and $10^7$, respectively (``{\it helicity
suppression}''). Although branching fractions at these levels are
beyond the reach of the Belle experiment with the current data sample,
an observation of either of the latter decay modes would be a clear
indication of physics beyond the SM. For example, the $\Btolv$ decay
rate may be enhanced in the minimal super-symmetric standard model
(MSSM) via intermediate charged Higgs bosons~\cite{ln_hou} or in the
Pati-Salam model of quark-lepton
unification~\cite{PatiSalam}. Similarly, $\Btolv$ may be mediated by
scalar super-symmetric particles in R-parity violating extensions of
the MSSM~\cite{BaekKim}. In this paper, we concentrate on the $\ell=e
$ and $\mu$ modes. The most stringent current upper limits for these
modes are $\BR (\Btoev) < 1.5\times 10^{-5}$~\cite{CLEOlv} and $\BR
(\Btomuv) < 2.1\times 10^{-5}$~\cite{babarlv} at the 90\% confidence
level.

It is natural to extend the $\Btolv$ searches to the corresponding
radiative modes. Because of the additional photon, helicity
suppression does not occur in these modes; hence $\Btolvg$ decays
are predicted to occur with rates comparable to or possibly larger
than $\Btomuv$ decay.  The $\Btolvg$ decay has been of theoretical
interest as a means of probing aspects of the strong and weak
interactions of a heavy quark system~\cite{BGW}-\cite{KPYan}. In
theory, there are two contributions to $\Btolvg $: the internal
Bremsstrahlung (IB) process and the structure-dependent (SD)
process~\cite{BGW}. In the IB process, a photon is emitted in
either the initial or final state.  The amplitude for this process
is, however, suppressed by both helicity conservation and the
electromagnetic coupling to the photon.  In the SD process, the
photon is produced in the transition of the spin-0 $B^+$ meson to
a spin-1 off-shell vector ($B^{*+}$) or axial-vector ($B'$) meson.
Helicity suppression is avoided in this spin exchange. In a recent
study by Korchemsky, Pirjol, and Yan (KPY)~\cite{KPYan}, the
predicted branching fraction is in the range $(2-5)\times
10^{-6}$.  Current upper limits on these modes are obtained by
CLEO: $\BR (\Btoevg) < 2.0\times 10^{-4}$ and $\BR (\Btomuvg) <
5.2\times 10^{-5}$~\cite{CLEOlvg}.

In this paper, we present a preliminary search for the leptonic
and radiative leptonic $B$ meson decays $\Btomuv$, $\Btoevg$, and
$\Btolvg$.  A brief description of the Belle detector is given in
the next section. The analysis procedure and results are described
in Section~\ref{Btomuv} for $\Btomuv$ decays and in
Section~\ref{Btolvg} for $\Btolvg$ decays.  Throughout this paper,
charge conjugation is implied unless stated otherwise.

\section{The Belle detector}
\label{detector}

The event sample that we analyze for this study corresponds to an
integrated luminosity of $ 140~\ifb $ accumulated at the $
\Upsilon (4S) $ resonance and recorded by the Belle detector at
the KEKB $\epem$ collider~\cite{KEKB}.  This sample contains 152
million $\BBbar$ pairs.  We also use an event sample of $15~\ifb$
integrated luminosity recorded $60~\MeV$ below $\Upsilon(4S)$
resonance for background determination.

The Belle detector is a large-solid-angle spectrometer based on a
1.5~T superconducting solenoid magnet.  Tracking and momentum
measurements of charged particles are done with a 3-layer
double-sided silicon vertex detector (SVD) and a central drift
chamber (CDC). Identification of charged hadrons is provided by a
combination of three measurements: specific ionization ($dE/dx$)
in the CDC, photon yield in the aerogel threshold Cerenkov
counters (ACC), and time-of-flight information from a cylindrical
array of 128 scintillation counters (TOF).  Photons are detected
in an electromagnetic calorimeter (ECL) system made of an array of
8736 CsI(Tl) crystals surrounding the TOF system.  Electrons are
identified based on the $dE/dx $ measurement in the CDC, the
response of the ACC, and the position, shape and energy of the
electromagnetic shower registered in the ECL.  Electron
identification efficiency is over 95\% in the momentum range of
this analysis; the fake rate is below 0.5\%.  Muons are identified
by a resistive plate chamber system (KLM) located within the
solenoid's external return yoke.  Muon identification efficiency
is approximately 90\% in the momentum range of this analysis; the
fake rate is approximately 1\%.  The Belle detector is described
in detail elsewhere\cite{BelleNIM}.

%----------------------------------------------

\section{Search for $\Btomuv$}
\label{Btomuv}

The main signature for the $\Btomuv$ decay is a single muon having
approximately $2.6~\GeVc$ momentum. The muon candidate is selected
from the charged tracks that are produced near the interaction
point (IP), and identified by its penetration depth and profile in
KLM.  We use only well-identified muons detected in the barrel
region of the Belle detector, where muon identification efficiency
and fake rate are well understood.  Muon candidates that are
consistent with being produced via a $J/\psi \to \mpmm$ decay are
vetoed.

To exploit the large missing energy and momentum of the undetected
neutrino in the $\Btomuv$ decay, we require that there should be
only one identified lepton in an event, and that the direction of
missing momentum be within the detector's fiducial volume ($|\cos
\theta_{\rm miss}| < 0.88$, where $\theta_{\rm miss}$ is the polar
angle of the missing momentum vector relative to the $z$ axis that
is opposite the positron beam line). These requirements make it
likely that the signal neutrino is the only undetected particle in
the event, so that the missing mass is consistent with zero.  We
require $-4.0<\Mm2 <4.3~(\GeVcc)^2$.

The  detected  particles  in  the  event other than the  signal
muon candidate are required  to be consistent with a $B$ decay
hypothesis.  This is tested using two kinematic variables,  $\Mbc
\equiv \sqrt{\Ebeam^2 - (\sum_i {\mathbf{p}}_i)^2}$ and $\DE
\equiv \sum_i E_i  - \Ebeam$ where $\Ebeam$ is  the  beam energy
measured in  the center-of-mass (CM) frame and  the summations are
over  all detected particles except the signal muon. We require
each event to lie within the region $5.1\,\GeVcc < \Mbc <
5.3\,\GeVcc$ and $-2.0\,\GeV < \DE < 1.5\,\GeV$.
The formation of this second $B$ meson's four-momentum permits us
to determine the reference frame in which the parent of the signal
muon is at rest (``$B$ rest frame'').  We reject events for which
the muon momentum in the $B$ rest frame lies outside the range
$2.58\,\GeVc <  \pmuB  < 2.8\,\GeVc$.

The dominant background arise from $\BtoXulnu$ decays and from the
continuum process $\epem \to \qqbar$ ($q=u,d,s,c$). There is some
additional contamination due to misidentification of a hadron as
the signal muon.

The lepton from $\BtoXulnu$ decays (but not the more abundant
$B\to X_c \ell \nu_\ell$ decays) can have sufficient energy to populate
the $\pmuB$ acceptance window. These events are suppressed by the
$\Mbc$--$\DE$ cuts, since the second $B$ meson inherits the
additional particles associated with the $X_u$ hadronic system.

Since a  $B$ meson pair  produced at the $\Upsilon(4S)$  resonance
has very  little kinetic  energy  in the  CM  frame ($\gamma\beta
\approx 0.06$), the event topology of $\BBbar $ events is
spherical while that of the continuum events, where the light
quarks are produced with large kinetic  energy, tends to  be
two-jet  like.   To suppress  continuum background,  we exploit
these event  shape differences  by combining several Fox-Wolfram
event shape moments~\cite{FoxWolfram} into a Fisher
discriminant~\cite{fisher} $\cal F$ whose coefficients are trained
using Monte Carlo (MC) simulated event samples to maximize the
distinction between signal and background. (See
Fig.~\ref{Fig:SFW}.) We also make use of the angle $\theta_T$
between the muon momentum and the thrust axis of the second $B$
meson in the CM frame. We suppress continuum background by
requiring ${\cal F} > 0.1$ and $\ACostt < 0.56$.

\begin{figure}[tb]
\begin{center}
\includegraphics[width=0.7\textwidth]{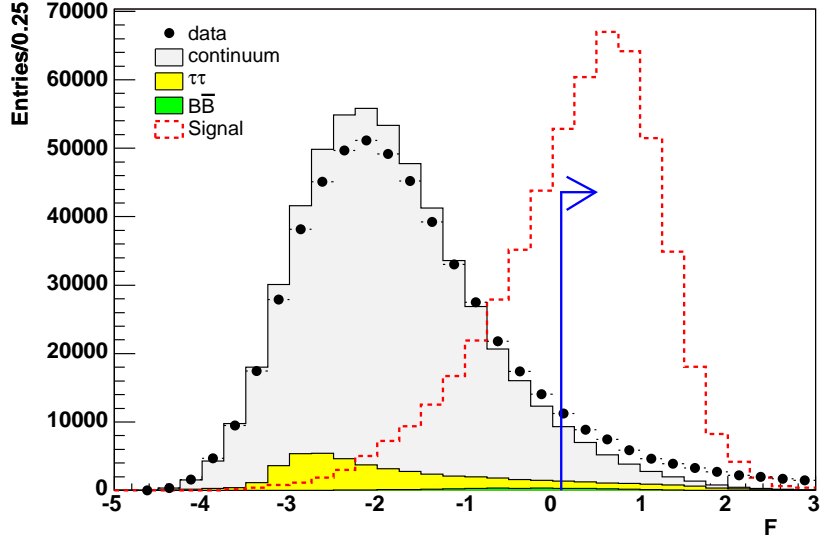}
\end{center}
\caption{Distributions of the event shape Fisher discriminant
  for signal and background MC samples. We require ${\cal F}>0.1$.}
\label{Fig:SFW}
\end{figure}

\begin{figure}[htb]
\begin{center}
\includegraphics[width=0.7\textwidth]
{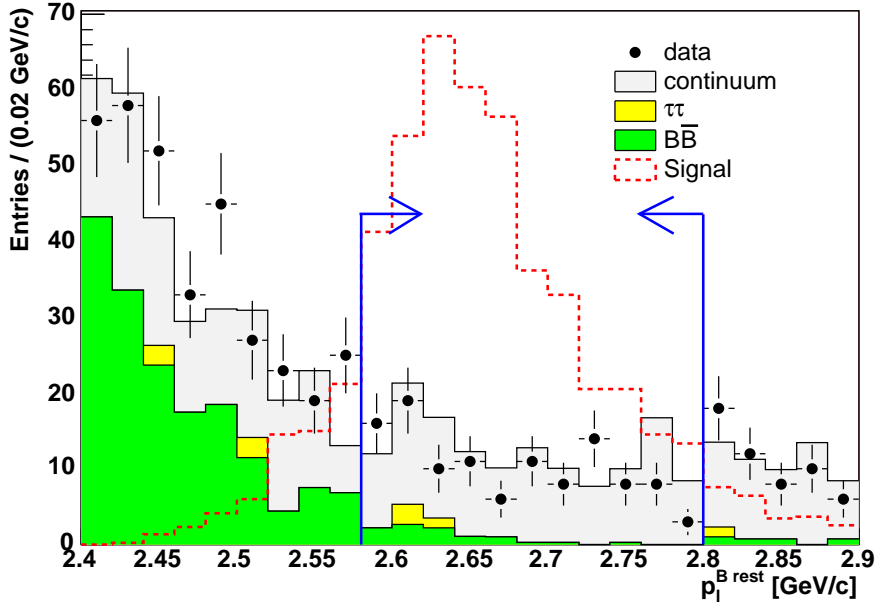}
\end{center}
\caption{Distributions of the muon momentum
  $\pmuB$ in the $B$ rest frame
  satisfying all selection criteria except the cut on $\pmuB$.
  We require $2.58\,\GeVc < \pmuB < 2.8\,\GeVc$.}
\label{pmuB_withcuts}
\end{figure}

Figure~\ref{pmuB_withcuts} shows the $\pmuB$ distributions for
signal and background MC events that satisfy all of the above
criteria except the $\pmuB$ cut. Signal yield and background
contamination are assessed in the two-dimensional subregion $\Mbc
> 5.26\,\GeVcc$ and $-0.8\,\GeV < \DE < 0.4\,\GeV$.
The definition of the signal region, as well as the requirements
on $\pmuB$, event-shape cuts and $\Mm2$ are optimized using signal
and background MC samples. Using a signal MC sample, the signal
detection efficiency is determined to be $(2.9\pm 0.1)\%$.

To estimate the background in the signal region, we define a
side-band region by $5.1 < \Mbc < 5.24~\GeVcc$ and $-2.0 < \DE <
1.5~\GeV$. This and the signal region are indicated in
Fig.~\ref{MBDE_muv}.  We fit the $\Mbc$ projection of the
side-band distribution of the MC background sample to an empirical
threshold function (``Argus function'')~\cite{Argus}. Then we
estimate the background in the signal region by fitting the $\Mbc$
projection of the data side-band with the same Argus function
where the fit parameters except for the overal normalization are
fixed to those obtained in the MC fit. The expected background in
the signal region, obtained by extrapolating the Argus function,
is $N_{\rm bkg} = 12.2_{-5.2}^{+5.4}$ events. Assuming no signal
events and  based on a Poisson fluctuation of background events,
we calculate mean expected upper limit as $1.7\times 10^{-6}$,
where backgrounds are subtracted.

\begin{figure}[tb]
\begin{center}
\begin{tabular}
{ @{\hspace{0.1cm}}c@{\hspace{1.0cm}}c@{\hspace{0.1cm}} }
\includegraphics[width=0.40\textwidth,height=0.25\textheight]
{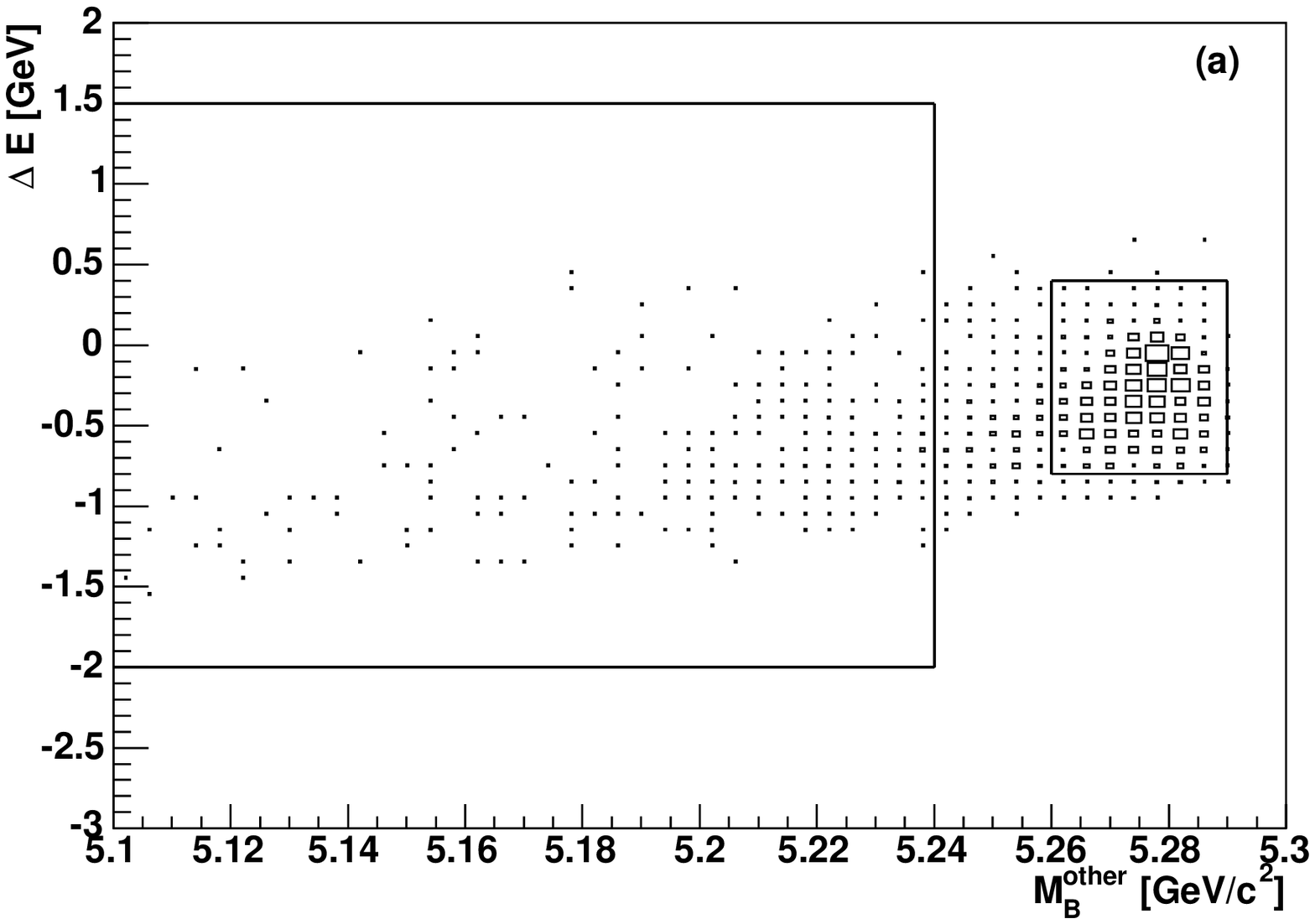}&
\includegraphics[width=0.40\textwidth,height=0.25\textheight]
{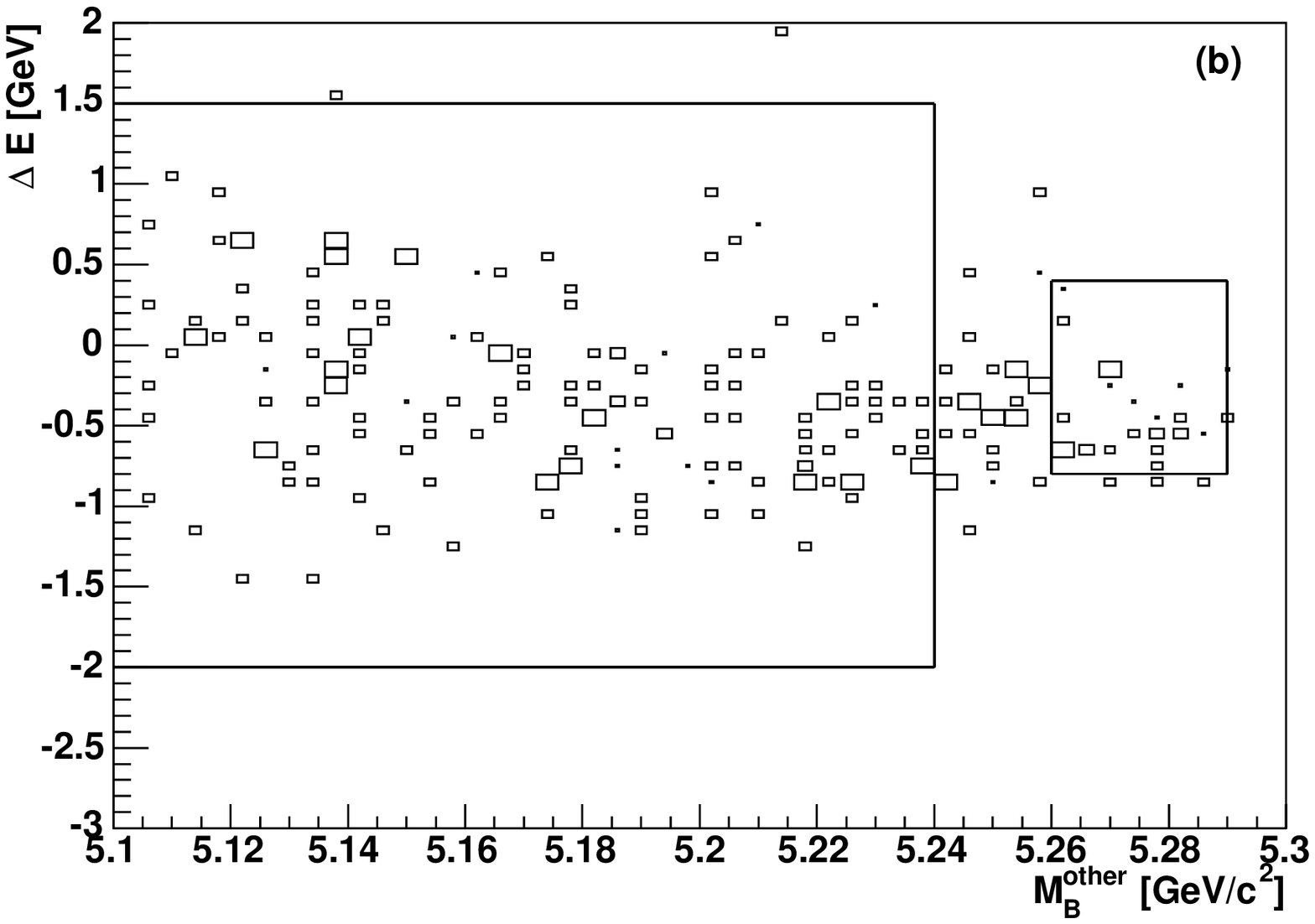}
\end{tabular}
\end{center}
\caption{Distributions in $\Mbc$-$\DE$ plane for (a) signal MC events and (b)
  background MC events satisfying all other event selection cuts. The signal
  region and side-band region are indicated with rectangles inside the
  plot.}
\label{MBDE_muv}
\end{figure}

Figure~\ref{Result_muv} shows the $\Mbc$ distributions for data,
background MC, and signal MC events after applying all selection
criteria except the $\Mbc$ cut. We find 4 events in the $\Mbc$ signal
region.  Without subtracting background, we convert this observation
and the above expectation for $N_{\rm bkg}$ into an upper limit of
7.99 on the signal yield at the 90\% confidence level.

\begin{figure}[tb]
\begin{center}
\medskip
\includegraphics[width=0.7\textwidth]{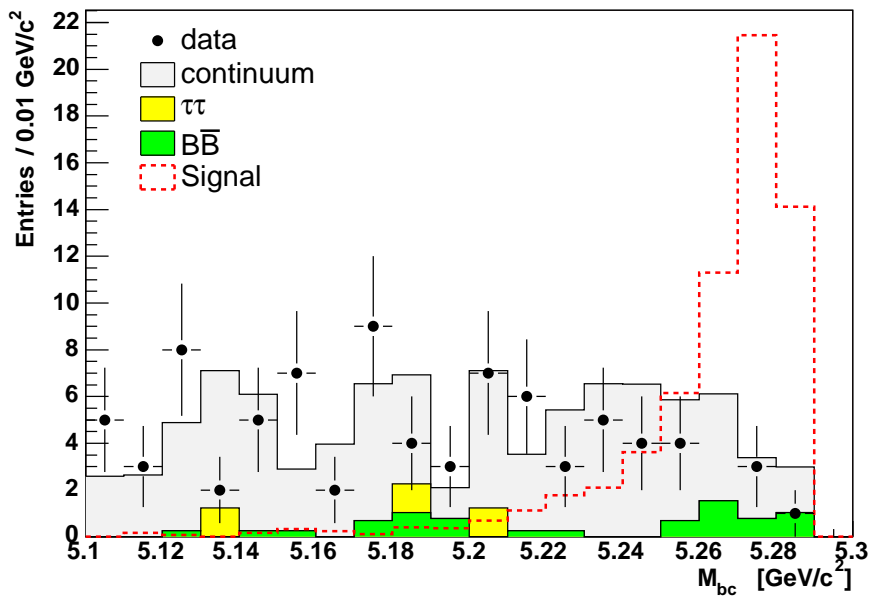}
\end{center}
\caption{The $M_{\rm bc}$ distributions for data (points
   with error bars) as well as signal and background MC events
   (histograms) for events satisfying all selection criteria except
   the cut on $M_{\rm bc}$. The signal region is $\Mbc > 5.26~\GeVcc$.}
\label{Result_muv}
\end{figure}

The systematic uncertainty in the $\Btolv$ analysis includes
uncertainties in the signal detection efficiency and in the background
estimation.  Since we do not subtract background in this analysis, we
consider only the uncertainty in the efficiency.

The uncertainty in $N_{\BBbar}$ is 0.3\%. The uncertainty in
track-finding efficiency is 2\% per track, and that in muon ID
efficiency is 2\%.  The effect of other event selection cuts is
studied by analyzing a calibration sample of fully reconstructed
$B^+ \to \bar{D}^{(*)0} \pip$ decays, where we treat the pion as a
signal muon and the $\bar{D}^{(*)0}$ as the accompanying neutrino.
We compare the efficiency of the event selection procedure between
data and MC to infer systematic uncertainty in the event selection
procedure.  The fractional difference of efficiency between data
and MC is 5.8\%.  Including the systematic uncertainty, the signal
efficiency is $(2.9\pm 0.2)\%$.

For a conservative upper limit, we reduce the signal efficiency by its
error and obtain the following upper limit for the $\Btomuv$ branching
fraction at 90\% CL:
$$\BR (\Btomuv) < \ResultBRBtomuv$$

\clearpage
%----------------------------------------------

\section{Search for $\Btolvg$}
\label{Btolvg}

We search also for the three-body radiative decays $\Btoevg$ and
$\Btomuvg$. The presence of the photon invalidates the
monochromaticity constraint on the lepton momentum in the $B$ rest
frame. Nevertheless, theoretical calculations\cite{BGW} predict
the lepton momentum spectrum to peak at the high end of its
kinematical range. While the photon's energy is typically much
softer than the lepton's, it exceeds that of most photons produced
at $\Upsilon(4S)$ resonance energy

%Event reconstruction begins by selecting the lepton-photon pair,
%and measuring the missing energy and momentum of the events. Then,
%we can reconstruct the signal decay candidates.

We select the primary muon or electron candidate with CM momentum
between $1.8$ and $2.8\,\GeVc$ using the particle identification
information described in section~\ref{detector}, and the primary
photon candidate as the most energetic of the neutral clusters in
the ECL (i.e., those not matched to any charged track) above
$1.0\,\GeV$ that does not appear to be affiliated with a $\piz$
meson. (A photon is considered to be a $\piz$ meson daughter if,
when combined with any other photon of energy $100\,\MeV$ or more,
it gives an invariant mass within $\pm 15\,\MeVcc$ of the nominal
$\piz$ mass.)  As before, we require that there be only one
identified lepton per event, that the direction of the missing
momentum be within the detector's fiducial volume, and that the
squared missing mass fall in the range $-2.0\,(\GeVcc)^2 < \Mm2 <
0.8\,(\GeVcc)^2$. In addition, we require the charge sum of all
detected particles be 0 or $\pm 1$.

As before, we calculate $\Mbc$ and $\DE$ for the remaining
particles in the event; here, candidates must satisfy
$5.11\,\GeVcc < \Mbc < 5.29\,\GeVcc$ and $-0.5\,\GeV < \DE <
0.125\,\GeV$. We also find that the lepton momentum ($\pLep$)
spectrum are different for different sources of backgrounds: the
lepton spectrum of $\BBbar$ MC events is peaking near $2.0~\GeVc$
and drops out rapidly beyond $\sim 2.3~\GeVc$ while that of the
continuum MC sample is much flatter than the others and reaches to
higher momentum regions ($\sim 3~\GeVc$ or beyond). The signal
yield is obtained by fitting the two-dimensional $\pLep$-$\Mbc$
distributions.

Continuum background is suppressed by a cut on the normalized
second Fox-Wolfram moment (rather than on the Fisher discriminant)
of $R_2 < 0.35$ for $\evg$ ($R_2 < 0.32$ for $\muvg$). The
relative fraction of each background component ($\BBbar$, $\Xulnu$
and continuum) are determined by fitting the lepton momentum distribution
within the $\Mbc$ side-band region of $5.11\,\GeVcc < \Mbc
< 5.20\,\GeVcc$ to the sum of the three empirically determined
shapes for these backgrounds.

For the $\Btomuvg$ mode, Fig.~\ref{CompareMm2} shows the
distributions within the signal region of the squared missing mass
and photon energy for data and combined background MC (with
relative weights equal to those measured in the $\Mbc$ side-band
region).  In both cases, the shapes of the data and combined
background are in good agreement. Similar agreement is seen for
the $\Btoevg$ mode.

\begin{figure}[tb]
\begin{center}
\begin{tabular}{cc}
\includegraphics[width=0.45\textwidth]{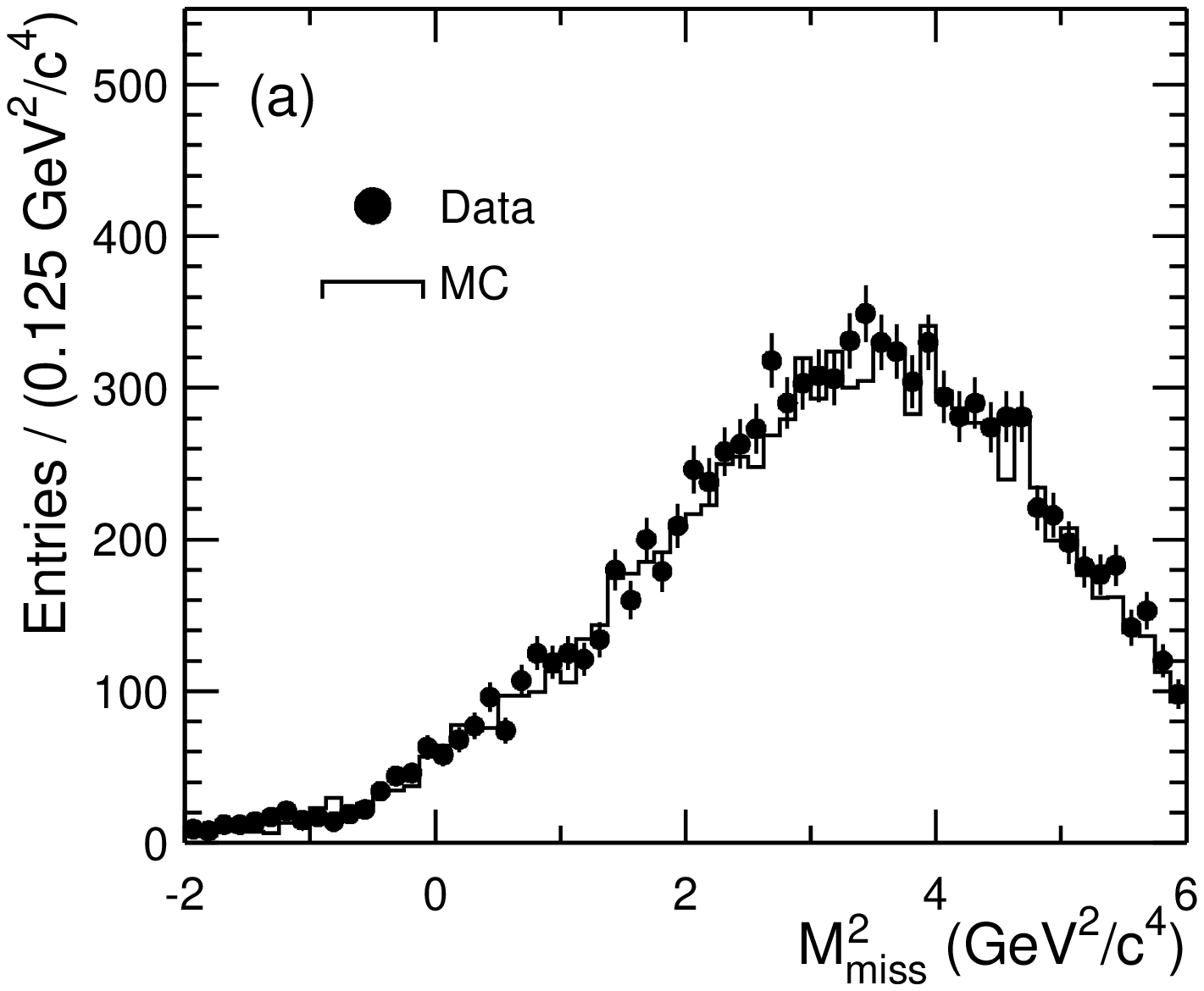}&
\includegraphics[width=0.45\textwidth]{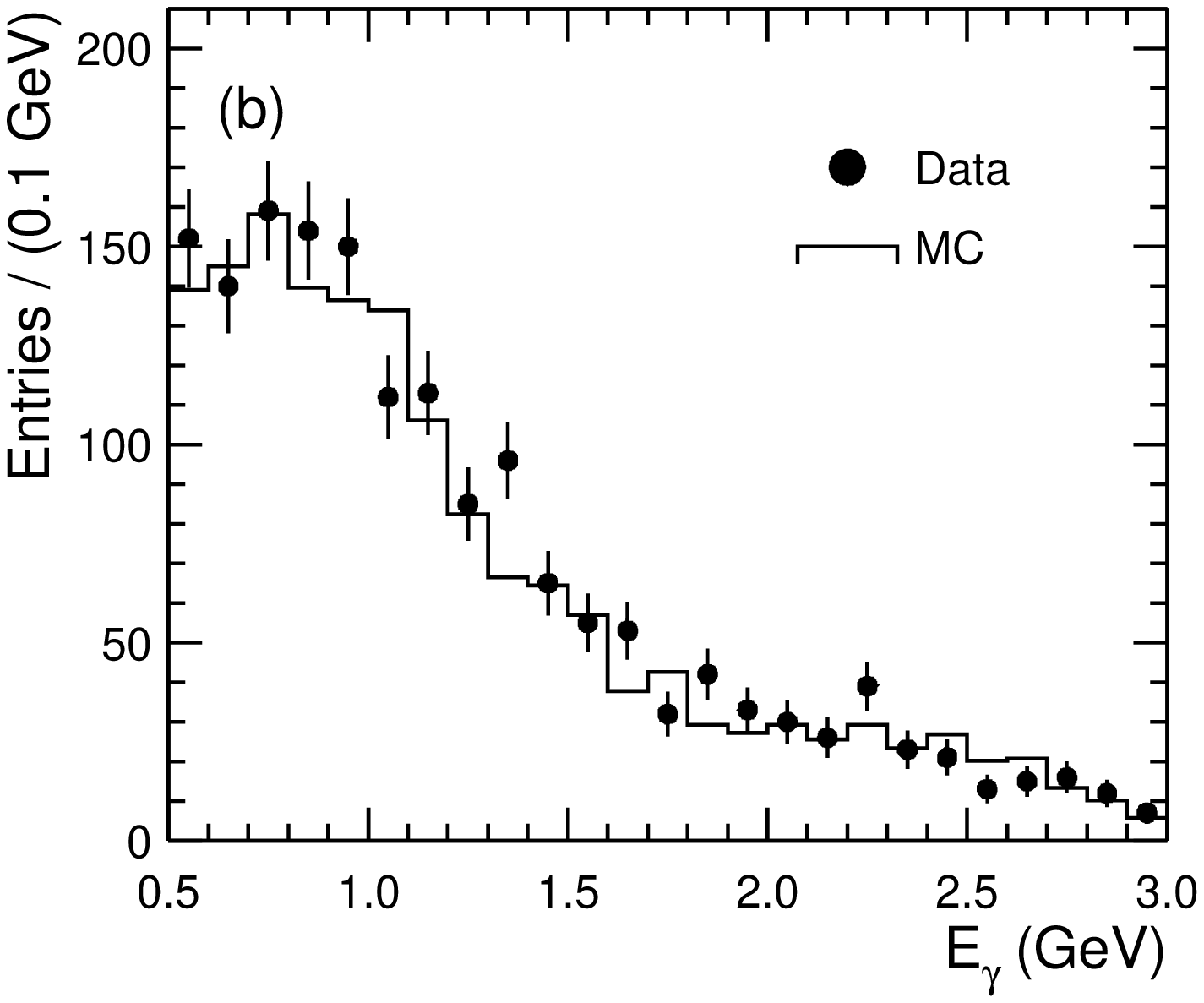}
\end{tabular}
\end{center}
\caption{\small Data and combined MC distributions of (a) squared
missing mass and (b) photon energy for the $\Btomuvg$ mode after
application of all event selection criteria except the cut on the
displayed quantity.}
\label{CompareMm2}
\end{figure}

Before fitting for the signal yield, the procedure is tested on an MC
sample made from a mixture of $\BBbar$, continuum and $B\to \Xulnu$
processes but without including any signal events. The signal yield
obtained are consistent with zero for both modes, and we take the
corresponding upper limits calculated without systematic uncertainty,
$1.4\times 10^{-5}$ for $\Btoevg$ mode and $1.0\times 10^{-5}$ for
$\Btomuvg$, as the search sensitivity for the corresponding mode.

We perform a binned maximum-likelihood fit to data in the
$\pLep$--$\Mbc$ plane; the two parameters in this fit are the
number of signal and combined-background events.
Figure~\ref{sigElFit} (Fig.~\ref{sigMuFit}) shows the projections
of the fit onto the $\Mm2$ and $\pLep$ axes for the $\Btoevg $
($\Btomuvg$) mode. The projection onto $\Mbc$ is made after
applying a cut $2.2\,\GeVc < \pLep  < 2.8\,\GeVc$, and the
projection onto $\pLep$ is made after applying a cut
$5.26\,\GeVcc < \Mbc < 5.29\,\GeVcc$.

Since neither mode shows a significant excess above the expected
background, we set upper limits by integrating the likelihood as a
function of the signal yield.  The dotted histograms in
Figs.~\ref{sigElFit} and \ref{sigMuFit} show the 90\% (statistical
error only) confidence level (CL) upper limits, added to the
background. The fitted yields after efficiency correction are
$970_{-550}^{+610}$ events in the $\Btoevg$ mode, and
$80_{-580}^{+660} $ events in the $\Btomuvg$ mode. Using the
statistical error only, the corresponding 90\% CL upper limits on
the branching fractions are $1.2\times 10^{-5}$ for $\Btoevg$, and
$0.7\times 10^{-5}$ for $\Btomuvg$.

Systematic uncertainties related to the signal efficiency are
estimated as in the $\Btomuv$ analysis. To be conservative, we
reduce the signal efficiency by the systematic error in
efficiency.

The systematic uncertainty in the yield extraction has three
categories: fitting method, background estimation, and signal
modelling. The uncertainty due to the fitting method is estimated
by repeating the fit under various scenarios. The systematic
effect of background uncertainty is estimated by repeating the fit
with each background component fraction changed by one standard
deviation. Similarly, the uncertainty due to signal modelling is
studied by varying the signal decay parameters in the signal MC
generation which is based on the KPY model~\cite{KPYan}. In each
category, we take the largest deviation from the default procedure
as an estimate of systematic uncertainty. Then the three errors
are added in quadrature for systematic uncertainty in the yield
extraction.

Assuming a Gaussian distribution for the systematic error, we
combine the systematic uncertainty in the yield extraction to the
statistical error by smearing the original likelihood function
obtained in the signal fitting with a Gaussian function based on
this error. Then we use this convolved likelihood function to
calculate the upper limits. The results, including the reduction of
efficiency by the corresponding systematic uncertainty, are:
\begin{eqnarray*}
\BR (\Btoevg ) & < & \ResultBRBtoevg \\
\BR (\Btomuvg) & < & \ResultBRBtomuvg~.
\end{eqnarray*}

\section{Summary}

In summary, we have searched for $\Btomuv$ and $\Btolvg$ decays
and found no significant excess in any mode. Therefore, we set
preliminary upper limits at the 90\% confidence level on the
branching fractions of $\BR (\Btomuv) < \ResultBRBtomuv$, $\BR
(\Btoevg) < \ResultBRBtoevg$ and $\BR (\Btomuvg) <
\ResultBRBtomuvg$. The result on $\Btomuv$ is more stringent than
the existing limit by a factor of three. The $\evg$ mode upper
limit is an improvement from the existing limit by an order of
magnitude and the limit on $\muvg$ is also improved by more than a
factor of two.

\section*{Acknowledgments}
%***** Acknowledgments *****
% Please paste this acknowledgement into your latex file.
%----------- Long version, for most papers -----------
We thank the KEKB group for the excellent operation of the
accelerator, the KEK Cryogenics group for the efficient operation of
the solenoid, and the KEK computer group and the National Institute of
Informatics for valuable computing and Super-SINET network support. We
acknowledge support from the Ministry of Education, Culture, Sports,
Science, and Technology of Japan and the Japan Society for the
Promotion of Science; the Australian Research Council and the
Australian Department of Education, Science and Training; the National
Science Foundation of China under contract No.~10175071; the
Department of Science and Technology of India; the BK21 program of the
Ministry of Education of Korea and the CHEP SRC program of the Korea
Science and Engineering Foundation; the Polish State Committee for
Scientific Research under contract No.~2P03B 01324; the Ministry of
Science and Technology of the Russian Federation; the Ministry of
Education, Science and Sport of the Republic of Slovenia; the National
Science Council and the Ministry of Education of Taiwan; and the U.S.\
Department of Energy.

\begin{figure}[tb]
\begin{center}
\includegraphics[width=0.9\textwidth]{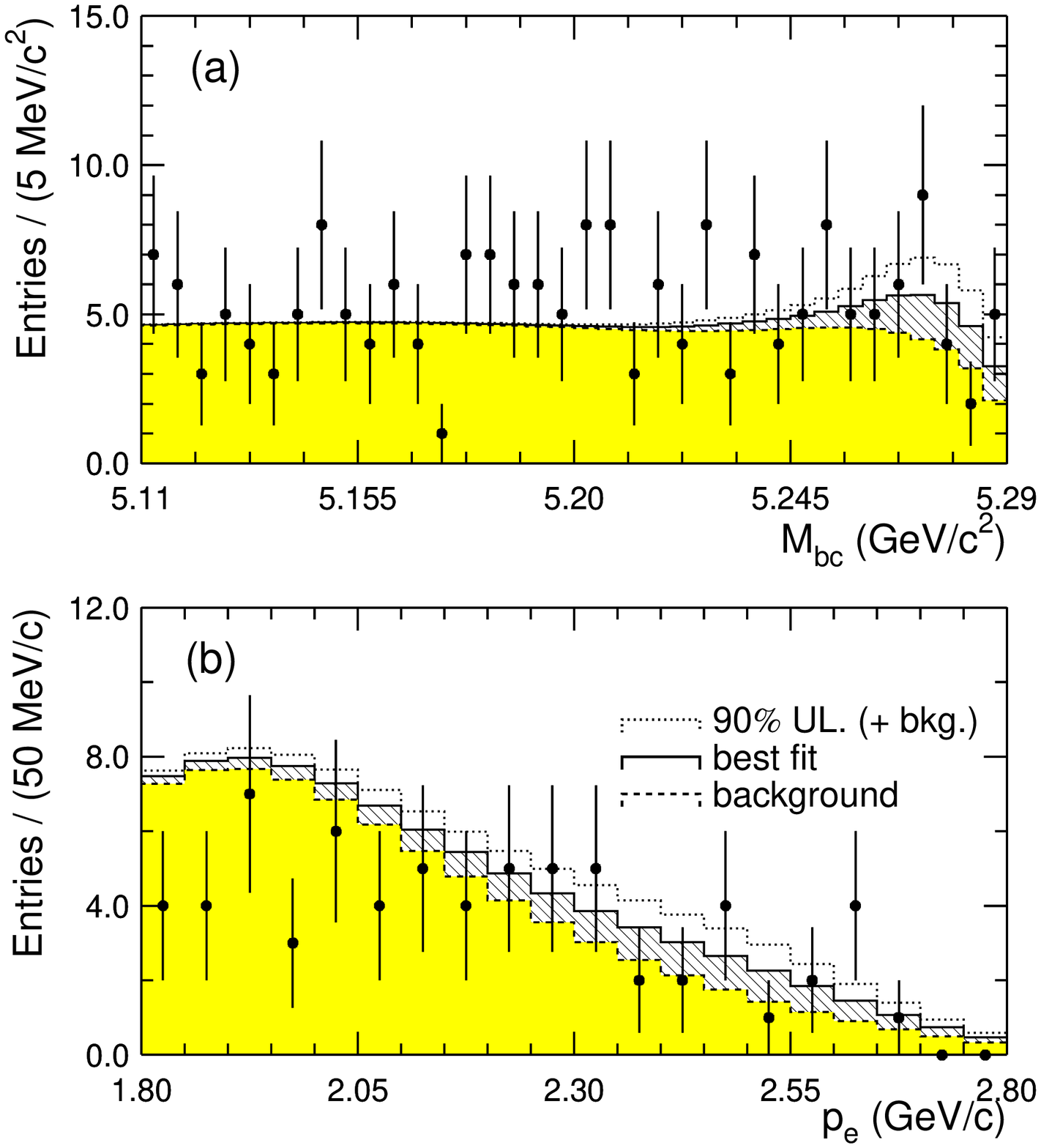}
\end{center}
\caption{\small Distributions of (a) beam constrained $B$ meson
mass and (b) electron momentum for events within the signal region
for the $\Btoevg$ mode. The projection onto $\Mbc$ is made after
applying a cut $2.2\,\GeVc < p_e  < 2.8\,\GeVc$, and the
projection onto $\pLep$ is made after applying a cut $5.26\,\GeVcc
< \Mbc < 5.29\,\GeVcc$. The points represent the data; the curves
represent the projections of the fit components.  Specifically,
the dashed curves show the background only; the solid curves show
the background and signal; the dotted curves show an alternate fit
with the signal component inflated to its 90\% CL upper limit.}
\label{sigElFit}
\end{figure}

\begin{figure}[tb]
\begin{center}
\includegraphics[width=0.9\textwidth]{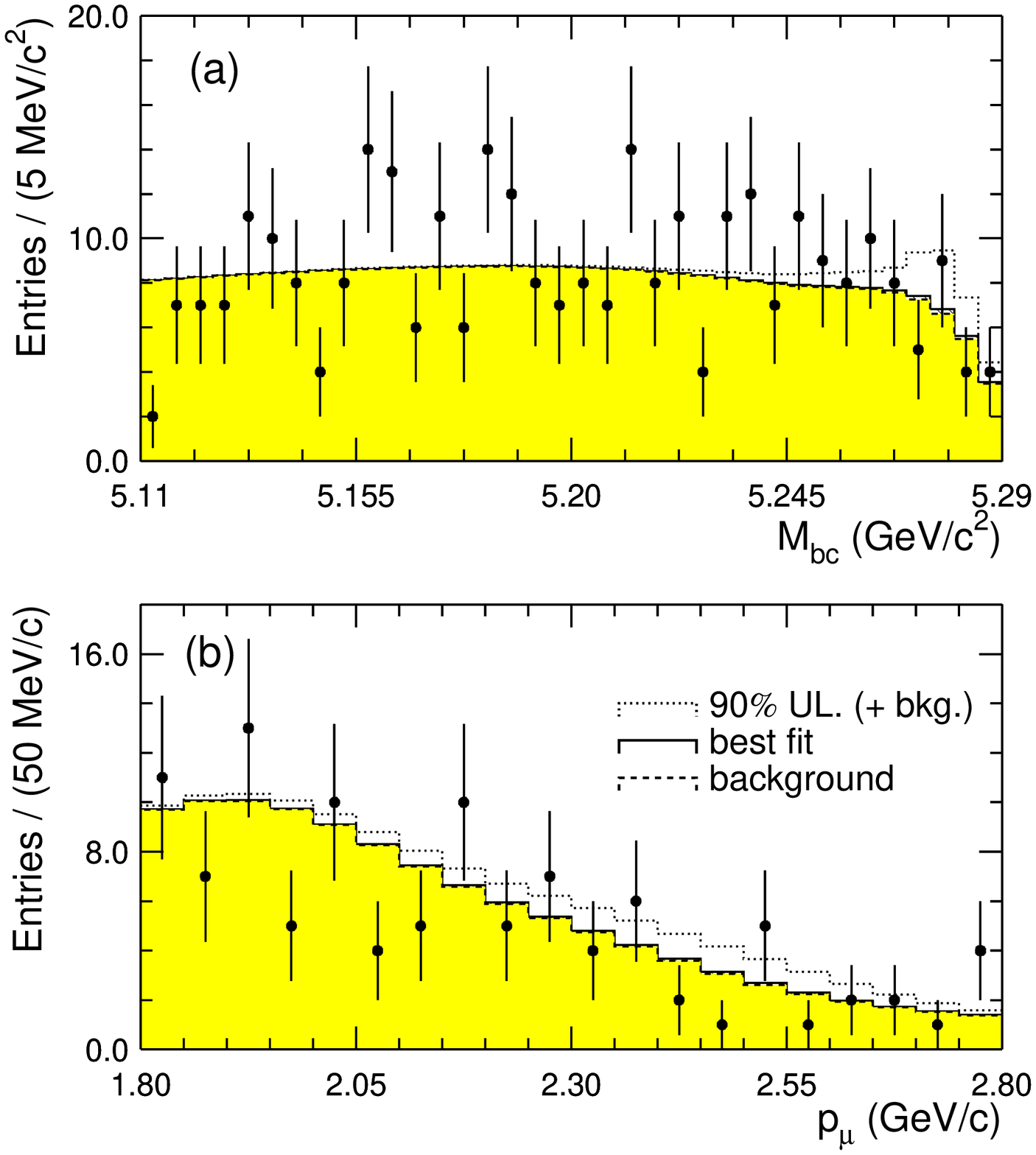}
\end{center}
\caption{\small Distributions of (a) beam constrained $B$ meson
mass and (b) electron momentum for events within the signal region
for the $\Btomuvg$ mode. The projection onto $\Mbc$ is made after
applying a cut $2.2\,\GeVc < p_\mu  < 2.8\,\GeVc$, and the
projection onto $\pLep$ is made after applying a cut $5.26\,\GeVcc
< \Mbc < 5.29\,\GeVcc$.  The points represent the data; the curves
represent the projections of the fit components.  Specifically,
the dashed curves show the background only; the solid curves show
the background and signal; the dotted curves show an alternate fit
with the signal component inflated to its 90\% CL upper limit.}
\label{sigMuFit}
\end{figure}

\end{document}